\pdfoutput=1
\documentclass[a4paper,10pt,twocolumn]{article}

\usepackage[english]{babel}
\usepackage[utf8]{inputenc}
\usepackage[T1]{fontenc}
\usepackage{amsmath,amsfonts,amssymb}
\usepackage{graphicx}
\usepackage{xcolor}
\usepackage{siunitx}
\usepackage{upgreek}
\usepackage{tabularx}
\usepackage{booktabs}
\usepackage{subfig}
\usepackage[labelfont=bf]{caption}[2004/07/16]

\usepackage[affil-it]{authblk}
\usepackage{abstract}

\usepackage{hyperref}
\hypersetup{colorlinks=true,urlcolor=oceanblue,citecolor=webgreen,linkcolor=blue,bookmarksopen=true,pdfauthor={M. Piazzi, J. Zemen, V. Basso},pdftitle={Ab-initio based analytical evaluation of entropy in magnetocaloric materials with first order phase transitions},pdfcreator={PDFLaTeX with hyperref package},pdfpagemode=UseOutlines,pdfpagelayout=SinglePage,pdfstartview=FitH}

\newcommand{\keywords}[1]{\textit{Keywords}: #1}
\newcommand{\PACS}[1]{\textit{PACS}: #1}

\newcommand{\beq}{\begin{equation}}
\newcommand{\eeq}{\end{equation}}
\newcommand{\ud}{\mathop{}\!\mathrm{d}}

\definecolor{oceanblue}{rgb}{0.125,0.125,0.9453125}
\definecolor{webgreen}{rgb}{0,0.3984375,0}

\begin{document}

\title{\textit{Ab-initio} based analytical evaluation of entropy in magnetocaloric materials with first order phase transitions}

\author[1]{M. Piazzi\thanks{Corresponding author.\\ \hspace*{0.55cm}E-mail address: \href{mailto:m.piazzi@inrim.it}{m.piazzi@inrim.it}}}
\author[2]{J. Zemen}
\author[1]{V. Basso}

\affil[1]{\small{Istituto Nazionale di Ricerca Metrologica, Strada delle Cacce 91, 10135 Torino, Italy}} 
\affil[2]{\small{Department of Physics, Blackett Laboratory, Imperial College London, London SW7 2AZ, UK}}

\date{}

\twocolumn[
\maketitle
\vspace{-1.1cm}
\begin{onecolabstract}
We combine spin polarised density functional theory and thermodynamic mean field theory to describe the phase transitions of antiperovskite manganese nitrides. We find that the inclusion of the localized spin contribution to the entropy, evaluated through mean field theory, lowers the transition temperatures. Furthermore, we show that the electronic entropy leads to first order phase transitions in agreement with experiments whereas the localized spin contribution adds second order character to the transition. We compare our predictions to available experimental data to assess the validity of the assumptions underpinning our multilevel modelling.

\medskip
\keywords{magnetocaloric effect, first order magnetic transitions, Density Functional Theory, Mn-based antiperovskites, electronic entropy}

\PACS{75.30.Sg, 75.30.Kz, 75.50.Ee, 71.20.Gj}  
\end{onecolabstract}]
\saythanks

\vspace*{0.2cm}

\section{Introduction}\label{sec:intro}
Promising materials for magnetic cooling applications, like La(Fe,Si)$_{13}$, MnFe($X$,P) (with $X =$ Ge, Si, As), FeRh and their related compounds, showing a giant magnetocaloric effect (MCE), are all characterized by first order magnetoelastic transitions. Moreover, the theory-led design of new compounds showing enhanced MCE is by no doubt an appealing task, hindered by the difficulty to predict phase transitions at finite temperatures from first principles. The development of an even approximate method allowing to predict this kind of transitions is therefore of high interest.

Spin polarised density functional theory (SDFT) can predict the ground state (GS) energy and the electronic density of states (DOS) of metallic compounds in ferro- (FM), antiferro- (AFM) or non magnetic (NM) states \cite{Bihlmayer-DFT}. The prediction of Curie points through SDFT has been also performed in the past \cite{Bihlmayer-DFT}. However, the presence of first order phase transitions and the evaluation of the entropy change at the transition are related to an estimate of the thermodynamic free energy associated with thermally disordered localized moments. Therefore, it is beyond the capabilities of standard SDFT. 

Moreover, the treatment of magnetostructural problems including lattice and magnetic degrees of freedom linked together has been formulated introducing microscopic Hamiltonians and this approach has been applied to magnetostructural transitions in martensites \cite{Comtesse-microham}. The method employed is based upon the Monte Carlo sampling of the microscopic energy states, so alternative analytical means would be desirable.

These observations led us to investigate thermodynamic models based on mean field theory (MFT) as an alternative approach. Indeed MFT allows for the analytical evaluation of magnetic entropy and for the inclusion of magnetoelastic effects and it is able to describe, within the approximation of the existence of magnetic sublattices, both the FM and AFM magnetically ordered states \cite{Piazzi-InverseMCE}. 

In this work we focus on a class of relevant magnetocaloric compounds, the Mn-based antiperovskites, i.e. $A$NMn$_3$ with $A=$ Ga, In, Ni, Sn \cite{Antipero-general}. These systems show a metallic behaviour with a low temperature (LT) magnetically ordered non collinear AFM state and a high temperature (HT) paramagnetic (PM) state. Many members of the $A$NMn$_3$ family show also a large magnetovolume effect \cite{Takenaka-antipero}. Together with the evaluation of the GS properties of these systems as a function of lattice distortion, it is interesting to foresee their transition temperature and entropy changes from the available SDFT data. Therefore, we have explored the possibility to combine SDFT zero temperature results with a MFT thermodynamic model to predict finite temperature properties.

In the paper we show that a MFT description of the non collinear AFM configuration of Mn-based antiperovskites is possible thanks to the particular arrangement of the magnetic moments placed at Mn sites within the unit cell. We use the results of SDFT calculations to set the parameters of the MFT model which includes magnetic, electronic and lattice contributions to the entropy and we derive the AFM-PM phase transition temperature $T_\text{t}$. We show that the transition temperature is critically dependent on the inclusion of the magnetic contribution to the total entropy and on the evaluation of the energy of the HT PM state. In particular, the inclusion of magnetic entropy is essential to lower $T_\text{t}$ as compared to the transition temperature based on the electronic entropy alone. However, the predicted $T_\text{t}$ is still higher than the experimental values. We discuss the possible origin of this discrepancy in the final section of this paper.  

\section{Antiperovskite systems}\label{sec:antipero_system}
We have investigated a class of magnetocaloric compounds, i.e. the antiperovskite manganese nitrides, having general formula $A$NMn$_3$, with $A$ representing a metal element, choosing $A$ = Ga, In, Ni, Sn. 

\begin{figure}[htbp]
\centering
\subfloat[][\label{Fig1a}]{\includegraphics[width=.38\textwidth]{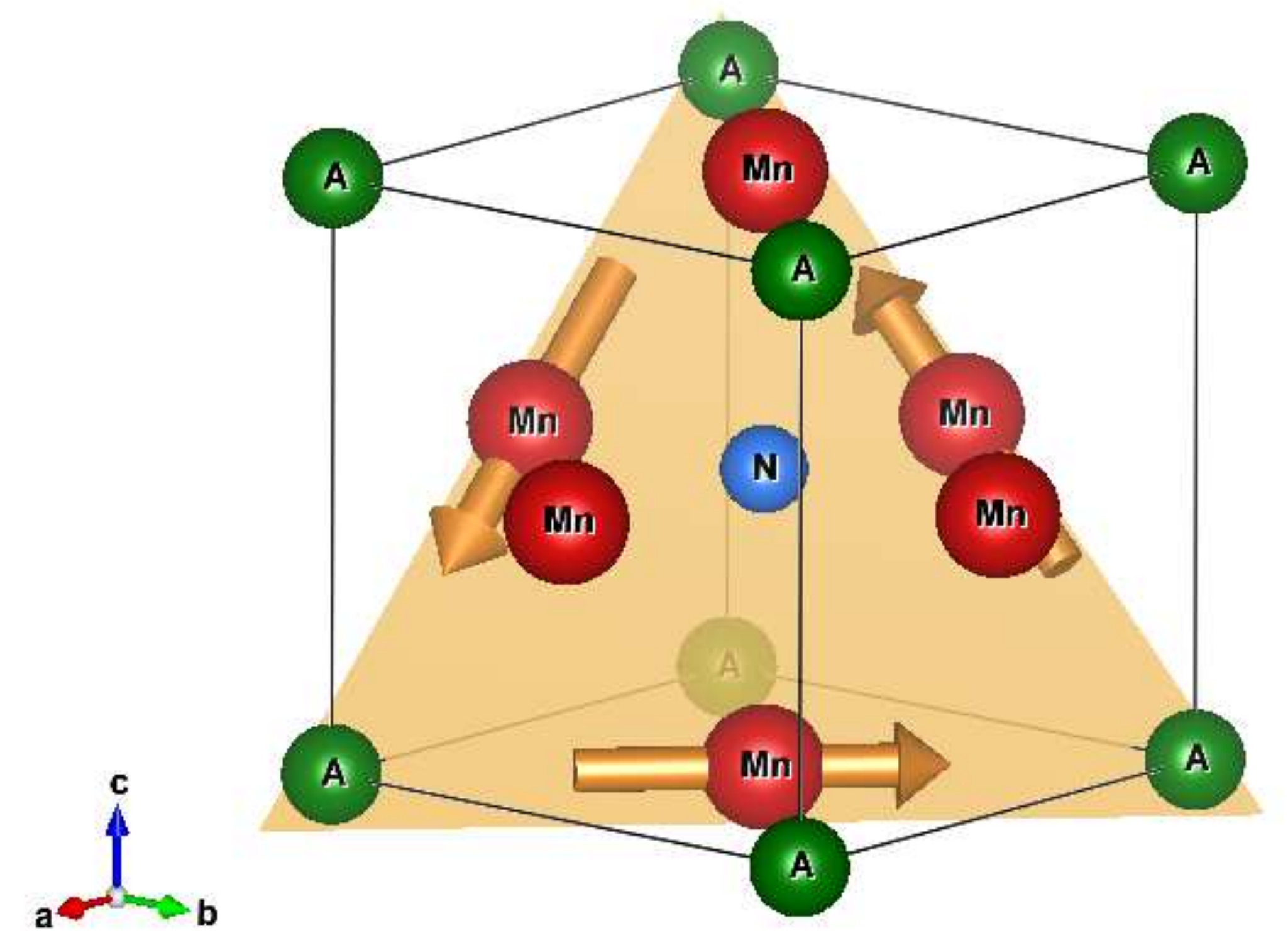}}\qquad\:
\subfloat[][\label{Fig1b}]{\includegraphics[width=.45\textwidth]{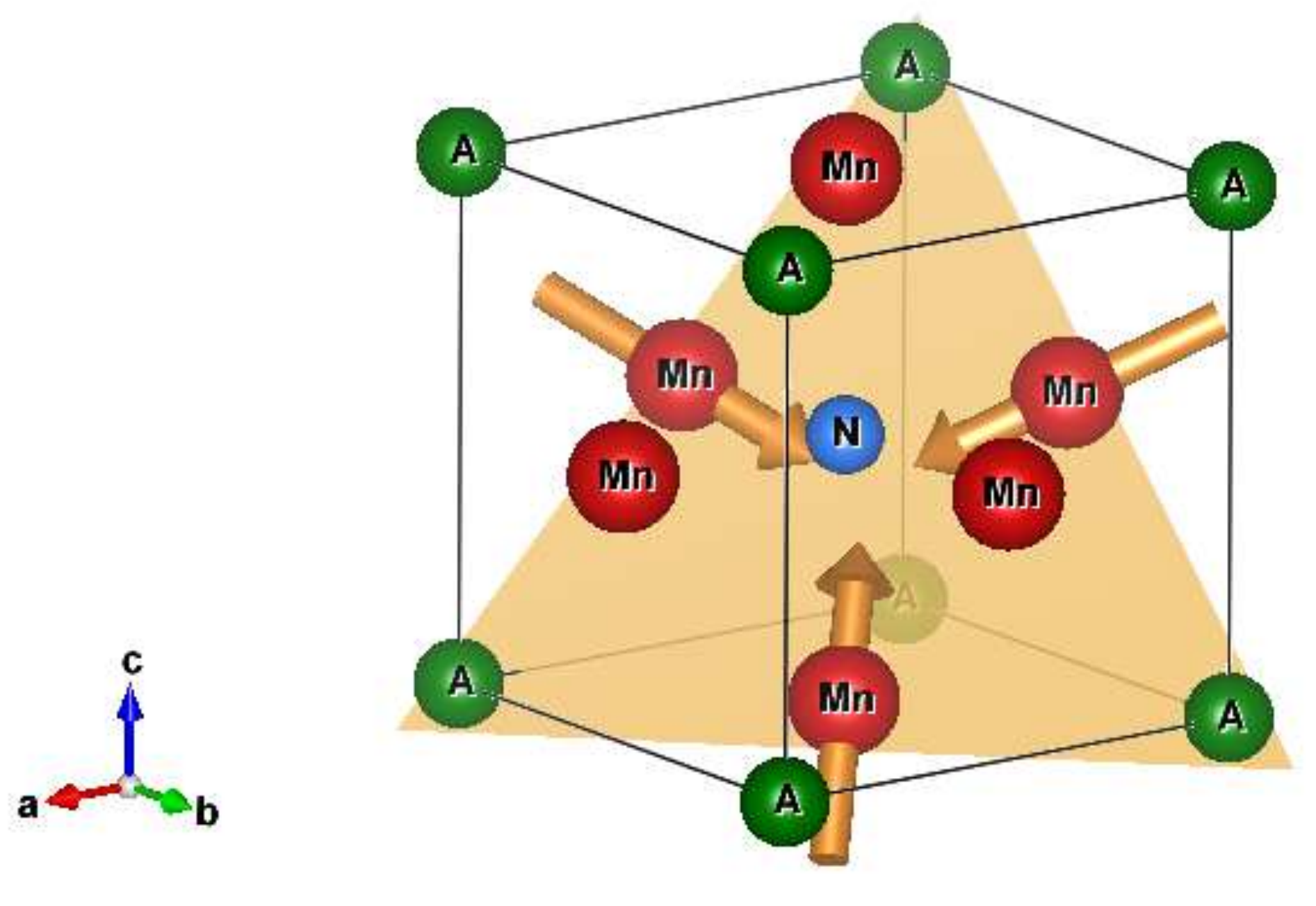}}
\caption{Representation of the antiperovskite nitride $A$NMn$_3$ unit cell ($A$ being a metal) for the (a) $\Gamma_{5\text{g}}$ and (b) $\Gamma_{4\text{g}}$ magnetic arrangements. (111) plane is highlighted; arrows indicate the nearest-neighbours Mn atoms' magnetic moments, laying in the plane. Drawings produced by VESTA software \cite{VESTA}.}
\label{Fig1}
\end{figure}

As already mentioned in Sec.~\ref{sec:intro}, these systems show a metallic behaviour with a LT non collinear AFM state and a HT PM state. The crystal has a cubic antiperovskite structure with five atoms per unit cell and space group symmetry \emph{Pm}$\bar{3}$\emph{m} \cite{Fruchart-Gaantipero,Antipero-general,Fruchart-Niantipero,Navarro-Gaantipero,Na-Niantipero,Sun-Snantipero,Sun-Inantipero,Antipero-review}. The three Mn atoms, placed at the centres of the cubic unit cell faces, are the magnetic ions in the compounds and they have magnetic moments $\boldsymbol{\mu}_i$ (with $i=A,B,C$), see \figurename~\ref{Fig1}. Therefore, we can distinguish three equivalent magnetic sublattices $A$, $B$, $C$, corresponding to different Mn atomic sites. In the non collinear AFM state the magnetic moments are arranged in the (111) plane either in the $\Gamma_{5\text{g}}$ triangular vortex structure (\figurename~\ref{Fig1a}) or in the $\Gamma_{4\text{g}}$ configuration (\figurename~\ref{Fig1b}) \cite{Antipero-general}. In both cases $\mu_A=\mu_B=\mu_C$ with $\mu_i=\lVert\boldsymbol{\mu}_i\rVert$ ($i=A,B,C$). SDFT simulations have shown that the difference between the $\Gamma_{5\text{g}}$ and $\Gamma_{4\text{g}}$ arrangements is introduced only by spin-orbit coupling and that the latter results to be a small contribution only. The GS arrangement depends on the particular $A$ metal element present in the compound. It is the $\Gamma_{5\text{g}}$ one for $A=$ Ga, In, while it is the $\Gamma_{4\text{g}}$ one for $A=$ Ni, Sn.  

In the unstrained case, the nearest-neighbours exchange energy of the system can be described by the Heisenberg Hamiltonian $\mathcal{H}=\sum_{\langle i,j\rangle} W_{ij} \boldsymbol{\mu}_i\cdot\boldsymbol{\mu}_j$, with $\langle i,j\rangle$ and $W_{ij}>0$ representing the $i$-th, $j$-th nearest-neighbours Mn sites and the exchange coupling parameters, respectively. We will assume the latter to be independent on the Mn atoms, so that $W_{ij}=W>0$ for all $i$, $j$. The sum in the Hamiltonian can be arranged so that it contains two different terms. The first term has the form $W/2\sum_\text{unit cells}{\lVert \boldsymbol{\mu}_A+\boldsymbol{\mu}_B+\boldsymbol{\mu}_C \rVert}^2$ and its minimization brings to the condition $\boldsymbol{\mu}_A+\boldsymbol{\mu}_B+\boldsymbol{\mu}_C=0$, corresponding to the non collinear AFM GS of the system. The second term is expressed as $-W/2\sum_\text{unit cells}\left(\mu_A^2+\mu_B^2+\mu_C^2\right)$. 

The MFT thermodynamic model is then developed by passing from the inclusion of the local magnetic moments $\boldsymbol{\mu}_i$ to the description of the macroscopic sublattice magnetizations (averaged magnetic moment per unit mass) $\boldsymbol{M}_i$ ($i=A,B,C$). It is reasonable to assume that the same two terms present in the Heisenberg Hamiltonian are also the terms appearing in the MFT magnetic free energy. Therefore, we can assume the first term of the free energy to be zero by taking the condition, valid also at finite temperature, $\boldsymbol{M}_A+\boldsymbol{M}_B+\boldsymbol{M}_C=0$. We are then left with a term $-W/2\sum_\text{unit cells}\left(M_A^2+M_B^2+M_C^2\right)$, with $M_i=\lVert \boldsymbol{M}_i\rVert$ ($i=A,B,C$), representing an effective FM-like interaction for the magnetizations of each sublattice. This way, the non collinear AFM order between $A$, $B$ and $C$ sublattices corresponds to a FM order in each sublattice. Since the strength of the interaction coupling $W$ and the exchange energy are the same for all of them, we can choose one sublattice as representative of the magnetic behaviour of the whole system, for example the $A$ one. Then, it is feasible to treat the system as a ferromagnet and to identify the order parameter appearing in the MFT free energy describing it (see Sec.~\ref{sec:thermod_model}) with the magnetization $M_A$ of the chosen sublattice. 

\section{Thermodynamic model}\label{sec:thermod_model}

\paragraph{Free energy and equation of state.}

The equation of state of a ferromagnet having magnetization $M_A$ in the scalar case in which a preferential axis coincident with the direction of the applied magnetic field has been fixed, neglecting magnetoelastic coupling and pressure effects, is given by:
\beq\label{eqstate}
\left.\frac{\partial G_\text{L}\left(T,M_A\right)}{\partial M_A}\right\rvert_T=0.
\eeq
$G_\text{L}\left(T,M_A\right)$ in Eq.~(\ref{eqstate}) is the Landau free energy mass density (in units \si{\joule\per\kilogram}) in the mean field approximation, expressed as
\beq\label{free_energy}
\begin{split}
G_\text{L}\left(T,M_A\right)&=G_0-\frac{1}{2}\mu_0 W M_A^2-T[S_\text{magn}\left(M_A\right)\\
&+S_\text{el}\left(T\right)+S_\text{latt}\left(T\right)]-\mu_0 H_A M_A.
\end{split}
\eeq
In Eq.~(\ref{free_energy}): $T$ and $\mu_0$ are the temperature and the vacuum permeability, respectively; $M_A$ represents the order parameter of MFT and in the antiperovskite system under investigation it is the magnetization per unit mass (in units \si{\ampere\square\metre\per\kilogram}) of the $A$ magnetic sublattice (see Sec.~\ref{sec:antipero_system} for further details on this system); $H_A$ is a fictitious applied magnetic field acting only on $M_A$, introduced for mathematical purposes but not linked directly to a physical magnetic field. 

The first term on the right hand side in Eq.~(\ref{free_energy}) is an overall constant setting the energy scale by changing the GS energy of the system. The second term represents the magnetic exchange energy, depending upon the dimensionful (\si{\kilogram\per\cubic\metre}) exchange coupling parameter $W>0$ introduced in Sec.~\ref{sec:antipero_system}. The third term is the sum of the magnetic ($S_\text{magn}$), electronic ($S_\text{el}$) and lattice ($S_\text{latt}$) contributions to the entropy per unit mass (i.e. in \si{\joule\per\kelvin\per\kilogram}) of the system. Finally, the last term resembles the Zeeman energy describing the interaction between the $A$ magnetic sublattice of the system and the fictitious magnetic field $H_A$.

It is more convenient to rewrite Eqs.~(\ref{eqstate}), (\ref{free_energy}) in terms of the dimensionless quantities $m_A=M_A/M_0$ ($-1\leq m_A\leq 1$), $h_A=H_A/\left(W M_0\right)$ and $t=T/T_0$ representing the reduced magnetization, magnetic field and temperature respectively, as follows:
\beq\label{eqstate_nodim}
\left.\frac{\partial g_\text{L}\left(t,m_A\right)}{\partial m_A}\right\rvert_t=0
\eeq
with $g_\text{L}\left(t,m_A\right)=G_\text{L}\left(t,m_A\right)/\left(\mu_0 W M_0^2\right)$, and
\beq\label{free_en_nodim}
\begin{split}
&G_\text{L}\left(t,m_A\right)=G_0-\frac{\mu_0 W M_0^2}{2} m_A^2\\
&\quad\;-n_\text{magn}k_\text{B} T_0 t\left[s_\text{magn}\left(m_A\right)+s_\text{el}\left(t\right)+s_\text{latt}\left(t\right)\right]\\[4pt]
&\qquad\qquad-\left(\mu_0 W M_0^2\right) h_A m_A.
\end{split}
\eeq
In Eq.~(\ref{free_en_nodim}), $M_0$ represents the saturation magnetization of the $A$ sublattice along the field direction. $n_\text{magn}$ is the number of magnetic ions per unit mass. We observe that it can be determined as $n_\text{magn}=N_\text{magn}^\text{f.u.}/M^\text{u.c.}$, where $N_\text{magn}^ \text{f.u.}$ and $M^\text{u.c.}$ are the number of magnetic ions per formula unit and the mass of the unit cell of the lattice depending on the system under investigation. $k_\text{B}$ is the Boltzmann constant. $T_0=a_J \mu_0 W M_0^ 2/\left(n_\text{magn}k_\text{B}\right)$ is the AFM-PM transition temperature at $h_A=0$ and in absence of electronic entropy; $a_J=\left(J+1\right)/\left(3 J\right)$, $J$ is the total angular momentum quantum number of magnetic atoms. Finally, in Eq.~(\ref{free_en_nodim}), we have introduced the dimensionless contributions to the entropy defined as $s=S/(n_\text{magn}k_\text{B})$.
  
It is worth noting that at a given $t$, the exchange and entropic energy terms of Eq.~(\ref{free_en_nodim}) behave differently as a function of magnetization, with the decrease in exchange energy at higher $m_A$ counteracting the increase in $-s_\text{magn}$. The stable magnetic state at any temperature $t$ and for various fields $h_A$ is determined, through Eq.~(\ref{eqstate_nodim}), by minimizing the total energy $g_\text{L}$. Therefore, the key point is to establish how each term appearing in Eq.~(\ref{free_en_nodim}) behaves at different temperatures.

\paragraph{Lattice entropy.}

The lattice contribution has been included for completeness by using the Debye model. Although magnetoelastic interactions exist and may be relevant \cite{Basso-MCE,Gruner-MCE,Jia-2006,Piazzi-conf} in the present paper we will not consider them. Therefore the entropy of the lattice gives a contribution which is only relevant for the specific heat values. 

In the Debye approximation the specific heat is then given by \cite{Mermin-book}:
\beq\label{latt_specheat}
c_\text{V}\left(t\right)=9 k_\text{B} n_\text{ions}{\left(\frac{t}{t_\text{D}}\right)}^3\int_0^{t_\text{D}/t} \frac{x^4 e^x}{{\left(e^x-1\right)}^2}\ud x
\eeq
where $n_\text{ions}=N_\text{ions}^\text{f.u.}/M^\text{u.c}$ is the mass density of lattice ions, $N_\text{ions}^\text{f.u.}$ being the number of lattice ions per formula unit, and $t_\text{D}=T_\text{D}/T_0$, $T_\text{D}$ being the Debye temperature. The lattice entropy, not depending on $m_A$, is then obtained from Eq.~(\ref{latt_specheat}) simply as $S_\text{latt}\left(t\right)=\int_0^ t c_\text{V}\left(t'\right)/t' \ud t'$, so that in dimensionless form it reads:
\beq\label{Slatt_nodim}
\begin{split}
&s_\text{latt}\left(t\right)\!=\!9\frac{n_\text{ions}}{n_\text{magn}} \int_0^ t \frac{\ud t'}{t'}\left[{\left(\frac{t'}{t_\text{D}}\right)}^ 3\!\!\!\int_0^{t_\text{D}/t'}\!\!\!\!\!\!\frac{x^4 e^x}{{\left(e^x-1\right)}^2}\ud x\right]\\[0.2cm]
&\!\!=\!\!-3\frac{n_\text{ions}}{n_\text{magn}}\!\left[\ln{\left(\!1\!-\!e^{-t_\text{D}/t}\right)}\!\!-\!4{\left(\frac{t}{t_\text{D}}\right)}^3\!\!\!\!\int_0^{t_\text{D}/t}\!\!\!\!\!\!\frac{x^ 3}{e^ x-1} \ud x\right]\!\!.
\end{split}
\eeq

\paragraph{Electronic entropy.} 
 
$S_\text{el}$ describes the contribution to the entropy per unit mass due to the conduction electrons present in a metallic compound and it is given by $S_\text{el}\left(T\right)=\left(\gamma/\rho_\text{el}\right)T$ or, in dimensionless form, by:
\beq\label{Sel_nodim}
s_\text{el}\left(t\right)=\frac{T_0}{n_\text{magn}k_\text{B}} \frac{\gamma}{\rho_\text{el}} t.
\eeq 
In Eq.~(\ref{Sel_nodim}) $\rho_\text{el}$ is the conduction electrons' mass density (in units \si{\kilogram\per\cubic\metre}) and $\gamma={\left(\uppi k_\text{B}\right)}^2 \mathcal{D}(E_\text{F})/3$ (see for example \cite{Mermin-book}) is the Sommerfeld coefficient depending on the electronic DOS at the Fermi level, i.e. on $\mathcal{D}(E_\text{F})$, $E_\text{F}$ being the Fermi energy. 

For non metallic systems the effect of this contribution is safely negligible. On the opposite, in metallic compounds $s_\text{el}$ may play an important role in determining the behaviour of the system close to the phase transition. Whenever the conduction electrons can be treated as a Fermi gas of non interacting particles, the DOS at $E_\text{F}$ is easily expressed in an analytical way \cite{Mermin-book} and the entropy takes the form $s_\text{el}\left(t\right)=\left(n_\text{el}/n_\text{magn}\right)\,\uppi^2 k_\text{B}T_0/\left(2 E_\text{F}\right)\,t$, with $n_\text{el}$ representing the number of conduction electrons per unit mass. 

However, the free electrons approximation is not usually suitable for the description of real materials and other methods are required to compute the DOS. Among them, we have used SDFT since it is one of the most powerful tool to reach this aim. Indeed, it can predict the $\gamma$ coefficient of a system having conduction electrons, as a metal, both in magnetically ordered or NM configurations. Moreover we have also assumed the coefficient to vary smoothly as a function of the magnetization $m_A$ in the following way:
\beq\label{gammacoeff}
\gamma\left(m_A\right)=\gamma_\text{NM}-m_A^2\Delta \gamma.
\eeq
$\Delta \gamma=\gamma_\text{NM}-\gamma_\text{M}$ and $\gamma_\text{NM}$, $\gamma_\text{M}$ are the Sommerfeld coefficients provided by SDFT at $m_A=0$ and $\lvert m_A\rvert=1$, respectively. The $m_A^2$ dependence of $\gamma$ has been chosen to preserve the time reversal symmetry of the magnetization, ensuring that $g_\text{L}$ must be an even function of $m_A$. With this assumption the $\gamma$ coefficient appearing in Eq.~(\ref{Sel_nodim}) is not a constant any more but it is replaced by Eq.~(\ref{gammacoeff}), making the electronic entropy a function also of $m_A$: $s_\text{el}=s_\text{el}\left(t,m_A\right)$. 
  
\paragraph{Magnetic entropy.}

Finally, concerning the magnetic contribution to the entropy, for localized magnetic moments systems statistical mechanics provides the following result, in which $s_\text{magn}$ is expressed as a function of $x_A=\mathcal{B}_J^{-1}\left(m_A\right)$ ($\mathcal{B}_J (x_A)$ being the Brillouin function) \cite{Coey-book}:
\beq\label{Smagn_nodim}
s_\text{magn}\left(x_A\right)=\ln{\left[\frac{\sinh{\left(\frac{2J+1}{2J}x_A\right)}}{\sinh{\left(\frac{x_A}{2J}\right)}}\right]}-x_A\mathcal{B}_J(x_A).
\eeq

\section{Results}\label{sec:results}

We have applied the multilevel approach introduced in Sec.~\ref{sec:intro}, combining SDTF and MFT results, to the antiperovskite compounds described in Sec.~\ref{sec:antipero_system}. On the one hand, since these systems have metallic behaviour, the MFT approach alone is not suitable to fully capture their thermodynamic properties. On the other hand, MFT allows the evaluation, at different $t$ and $h_A$, of the magnetic entropy $s_\text{magn}$ given by Eq.~(\ref{Smagn_nodim}). As we will show in Sec.~\ref{subsec:magnetiz_entropy}, this contribution is unavoidable to properly determine both qualitatively and quantitatively the character of the magnetic phase transition occurring in the system and its transition temperature $T_\text{t}$.    

\subsection{System parameters}\label{subsec:param}
The Vienna \textit{ab-initio} simulation package (VASP) code has been used to evaluate the energy and the DOS of the systems under investigation both in the NM and AFM configurations \cite{Kresse-VASP}. The DOS has been calculated using the Perdew-Burke-Ernzerhof generalized gradient approximation of the exchange correlation potential \cite{Perdew} and a $18\times 18\times 18$ $k$-point sampling. Moreover, same lattice parameters have been considered in evaluating the GS energies $E_\text{GS}^\text{NM}$, $E_\text{GS}^\text{AFM}$ for both the NM and AFM configurations. Therefore, the value obtained for the NM state may not necessarily be a GS energy but a good approximation of it. 
 
\tablename~\ref{Tab1} reports the values of the GS energy differences $E_\text{GS}^\text{NM}-E_\text{GS}^\text{AFM}$ between the NM and AFM states, and the GS energy $E_\text{GS}^\text{NM}=G_0$ corresponding to the NM configuration, as provided by VASP. The energy of the AFM state is representative of the FM state (i.e. $\lvert m_A\rvert=1$) energy, evaluated at $t=0$, in the thermodynamic model described by Eq.~(\ref{free_en_nodim}) (see Sec.~\ref{sec:thermod_model}). The energy of the NM configuration has been instead chosen as representative of the disordered PM state characterized by $m_A=0$, since in the NM state $M_A=M_B=M_C=0$. However, it is known that the PM and NM states may have different GS energies \cite{Staunton-2014}, as it will be discussed later (see Sec.~\ref{subsec:model}). With these choices, we can easily determine, through the SDFT results, the MFT parameters $\mu_0 W M_0^2=E_\text{GS}^\text{NM}-E_\text{GS}^\text{AFM}$ and $G_0=E_\text{GS}^\text{NM}$ (see \tablename~\ref{Tab1}). 

\begin{table*}[htbp]
\centering
\begin{tabularx}{.75\textwidth}{X*{4}{S[table-format=2.3]}}
\toprule
& {\textbf{GaNMn}$_{\boldsymbol{3}}$} & {\textbf{InNMn}$_{\boldsymbol{3}}$} & {\textbf{NiNMn}$_{\boldsymbol{3}}$} & {\textbf{SnNMn}$_{\boldsymbol{3}}$}\\
\midrule
$\boldsymbol{\Delta E_\text{\textbf{GS}}^\text{\textbf{NM-AFM}}}$  & 0.73 & 1.32 & 1.11 & 0.82\\
\textbf{[\si[detect-all=true]{\electronvolt}$/\text{f.u.}$]} & & & & \\
\midrule
$\boldsymbol{G_0=E_\text{\textbf{GS}}^\text{\textbf{NM}}}$ & -15.23 & -12.34 & -16.63 & -12.78\\
\textbf{[$\boldsymbol{\times}\!\!$ \SI[detect-all=true,per-mode=symbol]{e6}{\joule\per\kilogram}]} & & & & \\
\midrule
$\boldsymbol{\mu_0 W M_0^2}$ & 0.56 & 0.89 & 0.90 & 0.53\\
\textbf{[$\boldsymbol{\times}\!\!$ \SI[detect-all=true,per-mode=symbol]{e6}{\joule\per\kilogram}]} & & & & \\
\midrule
$\boldsymbol{\gamma_\text{\textbf{NM}}/\rho_\text{\textbf{el}}}$ & 0.097 & 0.107 & 0.073 & 0.096\\
\textbf{[\si[per-mode=symbol,detect-all=true]{\joule\per\kilogram\per\kelvin\squared}]} & & & & \\
\midrule
$\boldsymbol{\gamma_\text{\textbf{M}}/\rho_\text{\textbf{el}}}$ & 0.026 & 0.025 & 0.026 & 0.039\\
\textbf{[\si[per-mode=symbol,detect-all=true]{\joule\per\kilogram\per\kelvin\squared}]} & & & & \\
\bottomrule
\end{tabularx}
\caption{Values of GS energy differences $\Delta E_\text{GS}^\text{NM-AFM}=E_\text{GS}^\text{NM}-E_\text{GS}^\text{AFM}$, of parameters $G_0$, $\mu_0 W M_0^2$ of Eq.~(\ref{free_en_nodim}) and of $\gamma_\text{NM}/\rho_\text{el}$, $\gamma_\text{M}/\rho_\text{el}$ coefficients of Eq.~(\ref{gammacoeff}) for the antiperovskite compounds shown in the first row. Energies and $\gamma$ coefficients are provided by SDFT-VASP code for the NM and AFM configurations.}
\label{Tab1}
\end{table*}

\tablename~\ref{Tab1} shows also the values of the $\gamma$ coefficients for both the NM and AFM states, i.e. $\gamma_\text{NM}/\rho_\text{el}$ and $\gamma_\text{M}/\rho_\text{el}$, entering the electronic entropy Eq.~(\ref{Sel_nodim}). Since the latter is very sensitive to the numerical accuracy of the DOS, especially at LT, the $\gamma$ values here reported are still affected by an error which is below $10\%$. 

\subsection{Magnetization and entropy}\label{subsec:magnetiz_entropy}
After having evaluated the parameters present in Eq.~(\ref{free_en_nodim}) through SDFT, we have numerically solved Eq.~(\ref{eqstate_nodim}), searching for the global minimum of the Landau free energy. In metallic Mn-based antiperovskites, the magnetic moment is mainly due to the electron spin, so we have chosen $J=S=1/2$. This way we have determined $T_\text{t}$, $m_A$ and the total entropy $s_\text{tot}=s_\text{latt}+s_\text{el}+s_\text{magn}$ (with $s_\text{latt}$, $s_\text{el}$, $s_\text{magn}$ given by Eqs.~(\ref{Slatt_nodim}), (\ref{Sel_nodim}), (\ref{Smagn_nodim}), respectively) of the magnetic sublattice under investigation as a function of $t$ at various applied field $h_A$.

We have considered three cases, corresponding to the different contributions to the entropy that we have included into the free energy Eq.~(\ref{free_en_nodim}): the magnetic one $s_\text{magn}$ alone, the electronic one $s_\text{el}$ alone or both. This way it is possible to clarify how they affect the behaviour of the thermodynamic quantities describing the system.

\paragraph{Magnetization.}

\begin{figure}[htbp]
\centering
\includegraphics[width=.49\textwidth]{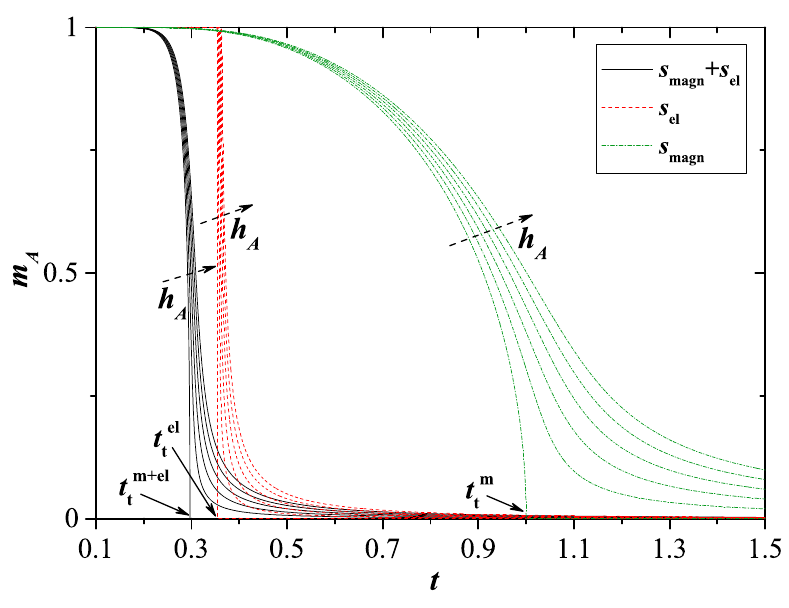}
\caption{$m_A$ vs. $t$ at $0\leq h_A\leq 0.05$ in steps of \num{0.01} and $J=1/2$ for different entropic contributions included in Eq.~(\ref{free_en_nodim}): \{$s_\text{el}$, $s_\text{magn}$\} (black solid lines), \{$s_\text{el}$\} (red dashed lines) and \{$s_\text{magn}$\} (green dash-dotted lines). Transition temperatures $t_\text{t}^\text{m+el}$,  $t_\text{t}^\text{el}$,  $t_\text{t}^\text{m}$ are defined in Eq.~(\ref{transtemp}). Remaining parameters $n_\text{magn}$, $\mu_0 W M_0^2$, $\Delta\gamma/\rho_\text{el}$ are set considering the values for GaNMn$_3$ compound (\tablename~\ref{Tab1}).}
\label{Fig2}
\end{figure}

The order of the magnetic phase transition and the transition temperature deeply rely on the various contributions to the entropy included in Eq.~(\ref{free_en_nodim}), as clearly shown by the magnetization curves reported in \figurename~\ref{Fig2}. At $h_A=0$, in all cases there is a phase transition between the LT-FM to the HT-PM state, as encompassed in the drop-off of $m_A$ at $t_\text{t}=T_\text{t}/T_0$. It is worth recalling that the LT-FM state in the chosen sublattice corresponds to the triangular AFM state in the whole antiperovskite lattice (see Sec.~\ref{sec:antipero_system}). The drop-off occurs suddenly and in a discontinuous way when considering only $s_\text{el}$, while it is smooth when only $s_\text{magn}$ or both the entropic contributions are included in Eq.~(\ref{free_en_nodim}). This fact means that in the former case the phase transition is first order but it slowly becomes more and more continuous by including spin entropy, being fully second order when $s_\text{magn}$ is the only contribution to the entropy. Moreover, the lowest $t_\text{t}$ value corresponds to the case in which both $s_\text{el}$ and $s_\text{magn}$ are non-zero, while it is enhanced if one of these two terms is neglected. The latter fact must be taken into account to avoid an overstimate of the transition temperature of a real system.

We can gather the same conclusions by looking at the analytical expression of $T_\text{t}$ at $h_A=0$. An approximate analytical expression for the transition temperature can be obtained evaluating explicitly Eq.~(\ref{eqstate_nodim}). Indeed, recalling that $\left.\partial s_\text{magn}(m_A)/\partial m_A\right\rvert_t=-\mathcal{B}_J^{-1}(m_A)$ and expanding the inverse Brillouin function at first order as $a_J \mathcal{B}_J^{-1}(m_A)\simeq m_A$, we end up with the following equation of state for $h_A=0$:
\begin{equation*}
\left.\frac{\partial g_\text{L}\left(t,m_A\right)}{\partial m_A}\right\rvert_t=c(t) m_A=0
\end{equation*}
with

\footnotesize
\beq\label{eqstate_firstorder}
c(t)\!=\!\!
\begin{cases} 
t-1 &\text { if } s_\text{el}=0\\[0.2cm] 
\frac{2\Delta \gamma}{\rho_\text{el}}\mu_0WM_0^2{\left(\frac{a_J}{n_\text{magn}k_\text{B}}\right)}^2 t^2-1 &\text{ if } s_\text{magn}=0\\[0.2cm] 
t\!-\!\!1\!+\!\frac{2\Delta \gamma}{\rho_\text{el}}\mu_0WM_0^2{\left(\frac{a_J}{n_\text{magn}k_\text{B}}\right)}^2\! t^2\!\!\!\!\! &\text{ if } s_\text{el},s_\text{magn}\neq 0 
\end{cases}
\eeq
\normalsize
where the parameters appearing in Eq.~(\ref{eqstate_firstorder}) have been introduced in Sec.~\ref{sec:thermod_model}. Since the transition temperature $t_\text{t}$ obeys $c(t_\text{t})=0$, we easily obtain:

\footnotesize
\beq\label{transtemp}
\begin{cases}
T_\text{t}^\text{m}=T_0=a_J\frac{\mu_0WM_0^2}{n_\text{magn}k_\text{B}} &\text{ if } s_\text{el}=0\\[0.2cm]
T_\text{t}^\text{el}=\sqrt{\rho_\text{el}\frac{\mu_0WM_0^2}{2\Delta \gamma}} &\text{ if } s_\text{magn}=0\\[0.2cm]
T_\text{t}^\text{m+el}\!\!=\!\frac{T_\text{t}^\text{el}}{2T_\text{t}^\text{m}}\!\left[\!\sqrt{4{\left(T_\text{t}^\text{m}\right)}^2\!+\!{\left(T_\text{t}^\text{el}\right)}^2}\!-\!T_\text{t}^\text{el}\right]\!\!\!\!\!\! &\text{ if } s_\text{el},s_\text{magn}\neq 0.
\end{cases}
\eeq
\normalsize
It is worth noting that for $\Delta\gamma\ll 1$, i.e. when $s_\text{el}$ is negligible and the dominant contribution to the entropy is due to the atomic spins, $T_\text{t}^\text{m+el}\rightarrow T_\text{t}^\text{m}$. In the opposite limit, $\Delta \gamma\gg 1$, when the electronic entropy drives the system through the transition, $T_\text{t}^\text{m+el}\rightarrow T_\text{t}^\text{el}$. Moreover, it is easily shown that $T_\text{t}^\text{m+el}<T_\text{t}^\text{el}$ and $T_\text{t}^\text{m+el}<T_\text{t}^\text{m}$: therefore, it is also analytically demonstrated that the combined action of magnetic and electronic entropies lowers the AFM-PM transition temperature in the antiperovskite compounds here considered. Finally, it is interesting to point out that $T_\text{t}^\text{el}<T_\text{t}^\text{m}$ only if $\Delta\gamma/\rho_\text{el}>{\left(n_\text{magn}k_\text{B}/a_J\right)}^2/\left(2\mu_0WM_0^2\right)$, so that $T_\text{t}^\text{m+el}< T_\text{t}^\text{el}<T_\text{t}^\text{m}$ only when $S_\text{el}\propto\Delta \gamma/\rho_\text{el}$ (see Eqs.~(\ref{Sel_nodim}), (\ref{gammacoeff})) is high enough with respect to the inverse of the exchange coupling energy. In particular, this is always true for the antiperovskite systems under investigation, as easily checked by looking at the values reported in \tablename~\ref{Tab1}.  

\paragraph{Entropy and entropy change.}

\begin{figure}[htbp]
\centering
\includegraphics[width=.49\textwidth]{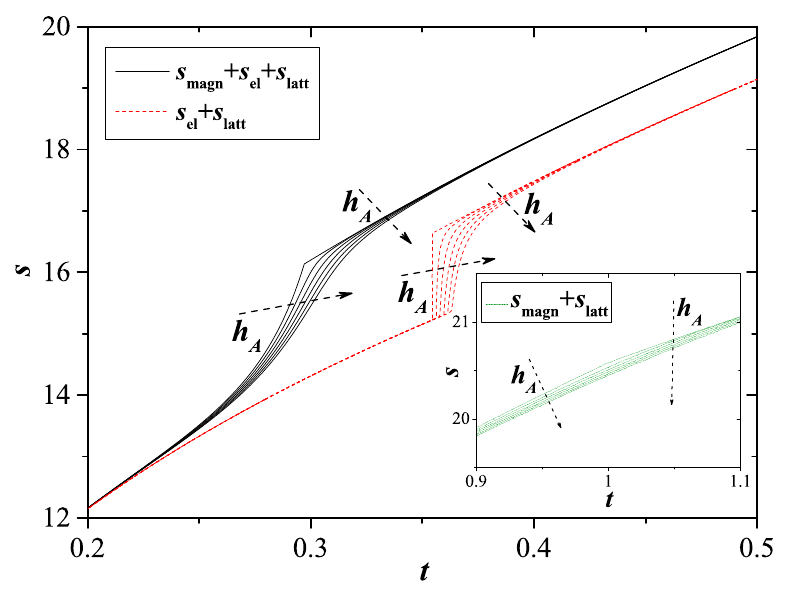}
\caption{Total entropy $s_\text{tot}$ vs. $t$ for the same fields $h_A$ and entropic contents set in \figurename~\ref{Fig2} (with $s_\text{latt}$ also included). Inset shows the case $s_\text{tot}=s_\text{latt}+s_\text{magn}$. The Debye temperature is $T_\text{D}=$ \SI{400}{\kelvin} \cite{Navarro-Gaantipero}. Remaining parameters  $J$, $n_\text{magn}$, $\mu_0 W M_0^2$, $\Delta\gamma/\rho_\text{el}$ are set as in \figurename~\ref{Fig2}. The meaning of the lines is the same of \figurename~\ref{Fig2}.}
\label{Fig3}
\end{figure}

\figurename~\ref{Fig3} shows the (dimensionless) entropy $s_\text{tot}$ for the same three entropic contributions considered for the magnetization curves. The entropy behaves qualitatively as the magnetization, thus having for $h_A=0$ a discontinuous jump at the first order AFM-PM transition when the spin entropy is neglected, becoming instead a smooth change typical of a second order transition when $s_\text{magn}$ is included. 

Finally, the entropy change occurring close to the phase transition, i.e. around $m_A=0$, depends upon the variation of the magnetic and electronic entropies. Expanding the magnetic contribution $s_\text{magn}$ given in Eq.~(\ref{Smagn_nodim}) around $x_A=0$ up to the second order in $m_A$, we obtain $s_\text{magn}\simeq \ln{\left(2J+1\right)}-m_A^2/\left(2a_J\right)$ . Adding the electronic entropy (see Eqs.~(\ref{Sel_nodim}), (\ref{gammacoeff})) we end up with a total entropy change close to the transition given by:
\beq\label{mce}
s_\text{tot}(m_A)-s_\text{tot}(0)=-m_A^2\left(\frac{1}{2 a_J}+\frac{\Delta\gamma}{\rho_\text{el}}\frac{T}{n_\text{magn}k_\text{B}}\right).
\eeq
It is clear from Eq.~(\ref{mce}) that the magnetic and electronic contributions to the entropy change may either sum or subtract depending on the sign of $\Delta\gamma$.

\subsection{Effective exchange coupling}\label{subsec:model}

Eq.~(\ref{transtemp}) links the transition temperature to the parameters $\Delta \gamma/\rho_\text{el}$ and $\mu_0 WM_0^2$. Substituting the values evaluated through SDFT for the Mn-based antiperovskite compounds shown in \tablename~\ref{Tab1}, we obtain the transition temperatures reported in \tablename~\ref{Tab2} which are pretty high.

\begin{table*}[htbp]
\centering
\begin{tabularx}{.7\textwidth}{>{\bfseries}X*{4}{S[table-format=4.0]}}
\toprule
 & {\textbf{GaNMn}$_{\boldsymbol{3}}$} & {\textbf{InNMn}$_{\boldsymbol{3}}$} & {\textbf{NiNMn}$_{\boldsymbol{3}}$} & {\textbf{SnNMn}$_{\boldsymbol{3}}$}\\
\midrule
$T_\text{t}^\text{m+el}$ [\si{\kelvin}] & 1670 & 2060 & 2590 & 1830\\ 
\midrule
$T_\text{t}^\text{el}$ [\si{\kelvin}] & 1995 & 2300 & 3100 & 2160\\
\bottomrule
\end{tabularx}
\caption{Transition temperatures for the antiperovskite compounds shown in the first row evaluated through Eq.~(\ref{transtemp}), in the case both $s_\text{magn}$, $s_\text{el}$ ($T_\text{t}^\text{m+el}$, second row) or $s_\text{el}$ alone ($T_\text{t}^\text{el}$, third row) are included in the free energy Eq.~(\ref{free_en_nodim}). The values of $\mu_0 WM_0^2$ and $\Delta\gamma/\rho_\text{el}$ for the various systems are reported in \tablename~\ref{Tab1}.}
\label{Tab2}
\end{table*}

First of all, we can clearly see that the inclusion of the magnetic entropy contribution in the free energy of the system (Eq.~(\ref{free_en_nodim})) lowers for all the compounds under investigation the transition temperature between the AFM and the NM states, as expected from Eq.~(\ref{transtemp}) (see also \figurename~\ref{Fig2}). However, the contribution of the spins is not enough to obtain reasonable and physically sound $T_\text{t}$ values. This result may be ascribed to the approximations we have employed in our model. On the one hand the use of the Heisenberg model to describe exchange interactions (see Sec.~\ref{sec:antipero_system}) may not be able to capture in detail the thermodynamic and magnetic behaviour in systems, such as the Mn-based antiperovskites, developing itinerant electron magnetism close to the transition temperature \cite{Takenaka-antipero}. On the other hand, we have estimated the exchange coupling coefficient $\mu_0WM_0^2$ for the compounds under investigation by considering the GS energies of the AFM and NM, instead of PM, states. It has been already shown in literature \cite{Lukashev-2008} that approximating the PM state with a collinear AFM configuration instead of a NM one lowers the AFM-NM energy difference, thus reducing also the value of $\mu_0WM_0^2$. For example, for the  GaNMn$_\text{3}$ compound, Lukashev et al. \cite{Lukashev-2008} have demonstrated that the GS energy difference $\Delta E_\text{GS}\simeq$ \SI{300}{\milli\electronvolt}/f.u., lower than the \SI{730}{\milli\electronvolt}/f.u. we have used here (see \tablename~\ref{Tab1}). Moreover, the coupling between lattice structure and exchange interactions, neglected in our model, may also contribute to the lowering of $\Delta E_\text{GS}$. In this sense it is worth mentioning that recently Gruner et al. \cite{Gruner-MCE} have shown that lattice entropy may act cooperatively with the magnetic contribution, thus promoting a lower phase transition temperature. The inclusion of magnetoelastic interaction will be the subject of future work.         

\begin{figure}[htbp]
\centering
\includegraphics[width=.49\textwidth]{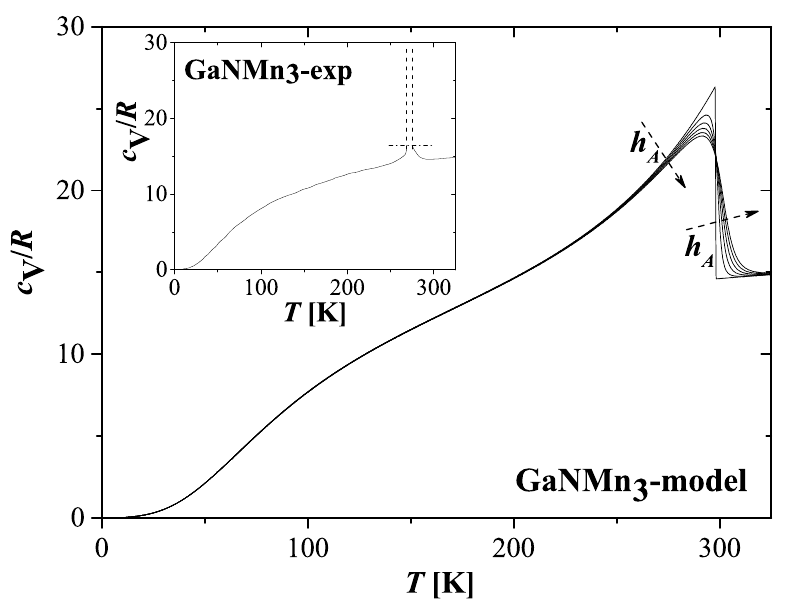}
\caption{Specific heat $c_\text{V}=T\partial S_\text{tot}/\partial T$, in units of $R=N_\text{A}k_\text{B}$ ($N_\text{A}$ being the Avogadro's number), as a function of $T$ for $0\leq h_A\leq 0.005$ in steps of $0.001$. Inset: experimental curve at zero applied magnetic field adapted from \cite{Navarro-Gaantipero}. Parameters set as:  $\left(\mu_0WM_0^2\right)^\text{eff}=$ \SI{4.25e4}{\joule\per\kilogram}; $J$, $T_\text{D}$, $n_\text{magn}$, $\Delta\gamma/\rho_\text{el}$ as in \figurename~\ref{Fig3} .}
\label{Fig4}
\end{figure}

We have then evaluated the specific heat $c_\text{V}=T \partial S_\text{tot}/\partial T$ for the GaNMn$_3$ compound. To obtain the resulting curve reported in \figurename~\ref{Fig4} we have used an effective exchange coupling coefficient $\left(\mu_0 WM_0^2\right)^\text{eff}=$ \SI{4.25e4}{\joule\per\kilogram}. The latter value has been chosen since it ensures that $T_\text{t}^\text{m+el}=$ \SI{298}{\kelvin}, i.e. the AFM-PM transition temperature experimentally known for GaNMn$_3$ \cite{Fruchart-Gaantipero}. Qualitatively similar behaviours are obtained for the other antiperovskite systems in \tablename~\ref{Tab1}.

The specific heat exhibits, at $h_A=0$, the $\lambda$-shaped peak characteristic of second order phase transitions. However, experiments lead to the observation of a first order phase transition characterized by a sharp peak at $T_\text{t}\simeq$ \SI{298}{\kelvin} for the GaNMn$_3$ compound (see Ref.~\cite{Fruchart-Gaantipero} and the inset in \figurename~\ref{Fig4}). The transition is accompanied by an abrupt change in the unit cell volume of the compound, although its cubic symmetry is preserved. The fact that magnetoelastic effects have been neglected in our model may explain the disagreement with the experimental results. 

\section{Conclusions}
We have used a multilevel approach, combining SDFT and MFT, to predict the magnetic transitions and thermodynamic properties of Mn-based antiperovskite compounds, a class of promising magnetocaloric materials. We have shown that the inclusion of spin entropy is important to lower the transition temperature of the systems. However, the fact that the presence of the magnetic entropy changes the transition from first order to second order and it does not lower enough the transition temperature, making it not comparable with experiments, means that several improvements can be done on the theory.  

On the one hand, an improvement in the estimate of the energy of the HT state of the system would probably bring to a more reasonable value of the MFT exchange coupling parameter, thus lowering the transition temperature. The approximation of the HT-PM phase with the NM state that we have employed in our SDFT calculations is probably too rough. Better estimates can be obtained by representing the PM phase as an AFM arrangement of the magnetic moments within each sublattice \cite{Lukashev-2008} as well as by using the disordered local moment (DLM) theory \cite{DLM-Gyorffy}. DLM is an alternative \textit{ab-initio} approach modelling the PM state at finite temperature as an ensemble of disordered non-zero local magnetic moments with random orientations. The validity of the DLM picture is guaranteed by the fact that in metals the kinetics governing moments orientations is much slower than the electronic motions one. It has been recently applied successfully to simulate AFM-FM transition in FeRh \cite{Staunton-2014}. 

On the other hand, the size of the overestimate in the transition temperature suggests that the itinerant electron aspect of the studied magnetic system is not negligible, so the Heisenberg treatment of the local moments may have limited applicability close to the transition temperature.  

Future developments of the present approach are therefore envisaged, as the better estimate of the GS energy, the improvement of MFT in its prediction of the spin contribution to the entropy and the inclusion of magnetoelastic effects to understand in detail the first order nature of the phase transitions occurring in several Mn-based antiperovskite compounds.
 
\section{Acknowledgements}
The research leading to these results has received funding from the European Community's \num{7}th Framework Programme FP\num{7}/\num{2007}--\num{2013} under Grant agreement no. \num{310748} ``DRREAM''.

\bibliographystyle{plain}
\bibliography{Piazzi_et_al_biblio}

\begin{thebibliography}{10}

\bibitem{Mermin-book}
N.~W. Ashcroft and N.~D. Mermin.
\newblock {\em Solid State Physics}.
\newblock Brooks/Cole - Cengage Learning, 1976.

\bibitem{Basso-MCE}
V.~Basso.
\newblock The magnetocaloric effect at the first-order magneto-elastic phase
  transition.
\newblock {\em J. Phys.: Condens. Matter}, \textbf{23}:226004, 2011.

\bibitem{Fruchart-Gaantipero}
E.~F. Bertaut, D.~Fruchart, , J.~P. Bouchaud, and R.~Fruchart.
\newblock Diffraction neutronique de {Mn$_3$GaN}.
\newblock {\em Solid State Commun.}, \textbf{6}:251, 1968.

\bibitem{Bihlmayer-DFT}
G.~Bihlmayer.
\newblock Density-functional {T}heory of {M}agnetism.
\newblock In H.~Kronmüller and S.~Parkin, editors, {\em Handbook of Magnetism
  and Advanced Magnetic Materials}, volume~1. J. Wiley \& Sons, Ltd., 2007.

\bibitem{Coey-book}
J.~M.~D. Coey.
\newblock {\em Magnetism and Magnetic Materials}.
\newblock Cambridge University Press, 2010.

\bibitem{Comtesse-microham}
D.~Comtesse, M.~E. Gruner, M.~Ogura, V.~V. Sokolovskiy, V.~D. Buchelnikov,
  A.~Grünebohm, R.~Arróyave, N.~Singh, T.~Gottschall, O.~Gutfleisch, V.~A.
  Chernenko, F.~Albertini, S.~Fähler, and P.~Entel.
\newblock First-principles calculation of the instability leading to giant
  inverse magnetocaloric effects.
\newblock {\em Phys. Rev. B}, \textbf{89}:184403, 2014.

\bibitem{Antipero-general}
D.~Fruchart and E.~F. Bertaut.
\newblock Magnetic {Studies} of the {Metallic} {Perovskite}-{Type} {Compounds}
  of {Manganese}.
\newblock {\em J. Phys. Soc. Jpn.}, \textbf{44}:781, 1978.

\bibitem{Fruchart-Niantipero}
D.~Fruchart, E.~F. Bertaut, R.~Madar, G.~Lorthioir, and R.~Fruchart.
\newblock Structure magnetique et rotation de spin de {Mn$_3$NiN}.
\newblock {\em Solid State Commun.}, \textbf{9}:1793, 1971.

\bibitem{Navarro-Gaantipero}
J.~Garc\'ia, J.~Bartolom\'e, D.~Gonz\'alez, R.~Navarro, and D.~Fruchart.
\newblock Thermophysical properties of the intermetallic {Mn$_3$MN} perovskites
  {II}. {H}eat capacity of manganese zinc nitride: {Mn$_3$ZnN} and manganese
  gallium nitride: {Mn$_3$GaN}.
\newblock {\em J. Chem. Thermodynamics}, \textbf{15}:1041, 1983.

\bibitem{Gruner-MCE}
M.~E. Gruner, W.~Keune, B.~Roldan Cuenya, C.~Weis, J.~Landers, S.~I. Makarov,
  D.~Klar, M.~Y. Hu, E.~E. Alp, J.~Zhao, M.~Krautz, O.~Gutfleisch, and
  H.~Wende.
\newblock Element-{Resolved} {Thermodynamics} of {Magnetocaloric}
  {LaFe}$_{13-x}${Si}$_x$.
\newblock {\em Phys. Rev. Lett.}, \textbf{114}:057202, 2015.

\bibitem{DLM-Gyorffy}
B.~L. Gyorffy, A.~J. Pindor, J.~Staunton, G.~M. Stocks, and H.~Winter.
\newblock A first-principles theory of ferromagnetic phase transitions in
  metals.
\newblock {\em J. Phys. F: Met. Phys.}, \textbf{15}:1337, 1985.

\bibitem{Jia-2006}
L.~Jia, G.~J. Liu, J.~R. Sun, H.~W. Zhang, F.~X. Hu, C.~Dong, G.~H. Rao, and
  B.~G. Shen.
\newblock Entropy changes associated with the first-order magnetic transition
  in {LaFe}$_{13-x}${Si}$_x$.
\newblock {\em J. Appl. Phys.}, \textbf{100}:123904, 2006.

\bibitem{Kresse-VASP}
G.~Kresse and D.~Joubert.
\newblock From ultrasoft pseudopotentials to the projected augmented-wave
  method.
\newblock {\em Phys. Rev. B}, \textbf{59}:1758, 1999.

\bibitem{Lukashev-2008}
P.~Lukashev, R.~F. Sabirianov, and K.~Belashchenko.
\newblock Theory of the piezomagnetic effect in {Mn}-based antiperovskites.
\newblock {\em Phys. Rev. B}, \textbf{78}:184414, 2008.

\bibitem{VESTA}
K.~Momma and F.~Izumi.
\newblock {VESTA} 3 for three-dimensional visualization of crystal, volumetric
  and morphology data.
\newblock {\em J. Appl. Crystallogr.}, \textbf{44}:1272, 2011.

\bibitem{Na-Niantipero}
Y.~Na, C.~Wang, L.~Chu, L.~Ding, J.~Yan, Y.~Xue, W.~Xie, and X.~Chen.
\newblock Preparation and properties of antiperovskite {Mn$_3$NiN} thin film.
\newblock {\em Mater. Lett.}, \textbf{65}:3447, 2011.

\bibitem{Perdew}
J.~P. Perdew, K.~Burke, and M.~Ernzerhof.
\newblock Generalized {Gradient Approximation Made Simple}.
\newblock {\em Phys. Rev. Lett.}, \textbf{77}:3865, 1996.

\bibitem{Piazzi-InverseMCE}
M.~Piazzi and V.~Basso.
\newblock Magnetocaloric effect at the exchange-inversion with magnetoelastic
  coupling.
\newblock {\em Physica B}, \textbf{473}:26, 2015.

\bibitem{Piazzi-conf}
M.~Piazzi, C.~Bennati, C.~Curcio, M.~Kuepferling, and V.~Basso.
\newblock Modeling specific heat and entropy change in {La(Fe-Mn-Si)$_{13}$-H}
  compounds.
\newblock {\em J. Magn. Magn. Mater.}, 2015.
\newblock doi:10.1016/j.jmmm.2015.07.055.

\bibitem{Staunton-2014}
J.~B. Staunton, R.~Banerjee, M.~dos Santos~Dias, A.~Deak, and L.~Szunyogh.
\newblock Fluctuating local moments, itinerant electrons, and the
  magnetocaloric effect: Compositional hypersensitivity of {FeRh}.
\newblock {\em Phys. Rev. B}, \textbf{89}:054427, 2014.

\bibitem{Sun-Snantipero}
Y.~Sun, Y.~Guo, Y.~Tsujimoto, C.~Wang, J.~Li, X.~Wang, H.~L. Feng, C.~I.
  Sathish, Y.~Matsushita, and K.~Yamaura.
\newblock Unusual magnetic hysteresis and the weakened transition behavior
  induced by {Sn} substitution in {Mn$_3$SbN}.
\newblock {\em J. Appl. Phys.}, \textbf{115}:043509, 2014.

\bibitem{Sun-Inantipero}
Y.~S. Sun, Y.~F. Guo, X.~X. Wang, W.~Yi, J.~J. Li, S.~B. Zhang, C.~I. Sathish,
  A.~A. Belik, and K.~Yamaura.
\newblock Magnetic and electrical properties of antiperovskite {Mn$_3$InN}
  synthesized by a high-pressure method.
\newblock {\em J. Phys.: Conf. Ser.}, \textbf{400}:032094, 2012.

\bibitem{Takenaka-antipero}
K.~Takenaka, M.~Ichigo, T.~Hamada, A.~Ozawa, T.~Shibayama, T.~Inagaki, and
  K.~Asano.
\newblock Magnetovolume effects in manganese nitrides with antiperovskite
  structure.
\newblock {\em Sci. Technol. Adv. Mater.}, \textbf{15}:015009, 2014.

\bibitem{Antipero-review}
P.~Tong, B.-S. Wang, and Y.-P. Sun.
\newblock Mn-based antiperovskite functional materials: {Review} of research.
\newblock {\em Chin. Phys. B}, \textbf{22}:067501, 2013.

\end{thebibliography}

\end{document}